\documentclass[pre,twocolumn,groupedaddress,showpacs]{revtex4}
\usepackage{graphicx}
\usepackage{dcolumn}
\usepackage{amsmath}
\usepackage{bm}

\begin{document}
\title{Computer simulation of topological defects around a colloidal 
particle or droplet dispersed in a nematic host}
\author{Denis Andrienko$^1$} 
\author{Guido Germano$^{1,2}$} 
\author{Michael P. Allen$^1$}
\affiliation{$^1$ H. H. Wills Physics Laboratory, University of Bristol, \\
Royal Fort, Tyndall Avenue, Bristol BS8 1TL, United Kingdom \\
$^2$ Theoretische Physik, Universit\"{a}t Bielefeld,
33501 Bielefeld, Germany}

\bibliographystyle{prsty}  
\begin{abstract}
  We use molecular dynamics to study the ordering of a nematic liquid crystal
  around a spherical particle or droplet.  Homeotropic boundary conditions and
  strong anchoring create a hedgehog director configuration on the particle
  surface and in its vicinity; this topological defect is cancelled by nearby
  defect structures in the surrounding liquid crystal, so as to give a uniform
  director field at large distances.  We observe three defect structures for
  different particle sizes: a quadrupolar one with a ring defect surrounding the
  particle in the equatorial plane; a dipolar one with a satellite defect at the
  north or south pole; and a transitional, non-equatorial, ring defect.  These
  observations are broadly consistent with the predictions of the simplest
  elastic theory.  By studying density and order-parameter maps, we are able to
  examine behaviour near the particle surface, and in the disclination core
  region, where the elastic theory is inapplicable.  Despite the relatively
  small scale of the inhomogeneities in our systems, the simple theory gives
  reasonably accurate predictions of the variation of defect position with
  particle size.
\end{abstract}
\maketitle
\section{Introduction}
\label{sec:intro}
Suspensions of solid particles or immiscible liquid droplets in a host fluid
(colloids) have attracted wide interest, due to a number of industrial
applications (they appear in food, paints, ink, drugs) and fundamental research
(they show features such as Casimir forces between the particles, formation of
supermolecular structures and novel phases).  Colloidal systems with a {\em
  liquid crystal} as a host fluid are of special
interest\cite{poulin.p:1997.a,poulin.p:1998.a,lev.bi:1999.a}. In the {\em
  nematic} liquid crystals considered here, while the system remains
translationally disordered and liquid-like, there is long-range ordering of
molecular orientations about a preferred direction, termed the {\em director}.
Elastic deformations of the director field around the colloidal particles
produce additional long-range forces between them. These interactions can be of
dipolar or quadrupolar type depending on the symmetry of the director field
around the particles \cite{lubensky.tc:1998.a}, and this in turn is extremely
sensitive to the {\em boundary conditions} on the particle surface and the size
of the particles
\cite{kuksenok.ov:1996.a,shiyanovskii.sv:1998.a,mondain-monval.o:1999.a}.

To understand the interaction between the particles, it is important to know the
liquid crystal ordering near one single particle.  This ordering is also
important for calculation of the elastic light scattering from nematic emulsions
\cite {lednei.m:1999.a}.  The complication here is the presence of {\em
topological defects} in such emulsions.  Isolated drops provide a spherical
confining geometry for the liquid crystal. Sufficiently strong homeotropic
anchoring of the director (that is, normal to the particle surface) then induces
a radial hedgehog defect with topological charge $+1$.  If the director field is
uniform far from the particle, i.e.\ the total charge of the whole system is
zero, topological considerations imply that an additional defect must be created
in the medium to compensate the radial hedgehog.

\begin{figure}
 \includegraphics[width=8cm]{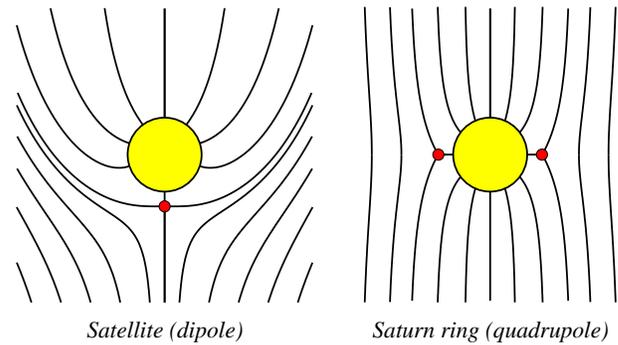}
 \caption[Sketches]{
 \label{fig:1} 
Sketches of the satellite defect and Saturn-ring defect.
}
\end{figure}

There are several types of defect which can arise in this case.  Two are
illustrated in Fig.~\ref{fig:1}. The first is a hyperbolic hedgehog with a
topological charge $-1$, called a dipolar or {\em satellite} defect.  The second
is a quadrupolar or {\em Saturn-ring} defect, i.e.\ a $-1/2$ strength
disclination ring that encircles the spherical particle.  Theoretical and
numerical work based on the elastic theory
\cite{ruhwandl.rw:1997.b,stark.h:1999} suggest that the dipole configuration is
stable for the micron-sized droplets which are usually realised experimentally;
the Saturn-ring configuration should appear if the droplet size is reduced and,
when present, it is always predicted to be most stable in the equatorial plane
normal to the director.  If the strength of the surface anchoring is weak
enough, a third, {\em surface-ring}, director configuration is possible
\cite{kuksenok.ov:1996.a,stark.h:1999}.  It has been also found that the {\em
  twisted dipole} configuration is possible for high enough ratios of twist to
splay elastic constants \cite{stark.h:1999}.

However, the elastic theory has several disadvantages because of the way it
treats the core structure of the defect embedded into the nematic host.  First,
the defect core region is considered as an isotropic inclusion with some unknown
free energy which is used as an adjustable parameter of the theory.  Second, the
liquid crystalline phase is treated as uniaxial.  However, near the defect core
(or indeed the particle surface) these assumptions are no longer valid. As has
already been demonstrated
\cite{biscari.p:1997.a,mottram.nj:1997.a,sigillo.i:1998.a,andrienko.d:2000.b},
the defect core of the simplest disclination line of strength $+1$ has a rather
complex structure: the order parameter varies near the core region; the uniaxial
phase becomes biaxial; even the density of the mesophase can be affected by the
presence of the wall or the defect \cite{andrienko.d:2000.b,allen.mp:1999.a}.

For a better description it is necessary to use the full order parameter tensor
to describe liquid crystalline ordering
\cite{sonnet.a:1995.a,schopohl.n:1987.a,schopohl.n:1988.a}.  This tensor
exhibits no anomalous behaviour at the singularity of the director field.
Moreover, one can resolve order parameter variation as well as biaxiality of the
defect core using a Landau-de~Gennes phenomenological expansion of the free
energy.  In this framework, only the liquid crystal density is assumed to be
fixed.  In its turn, density variation can be taken into account using Onsager
theory \cite{allen.mp:1999.a}, although that approach will not be pursued here.

Another approach to the study of defect core structure is to use computer
simulation. Indeed, computer simulation is a well established method to treat
bulk elastic coefficients \cite{allen.mp:1988.a,allen.mp:1990.a}, the surface
anchoring strength \cite{allen.mp:1999.a,andrienko.preprint}, and structures of
disclination cores \cite{andrienko.d:2000.b,hudson.sd:1993.a}.  This means that
computer simulation allows one to investigate the details of liquid crystalline
structure which cannot be resolved experimentally or using phenomenological
theories.

In this paper we present the results of molecular dynamics (MD) simulations of
the topological defects which appear in the liquid crystalline mesophase around
a spherical droplet or colloid particle.  Our study complements previous results
obtained by Billeter and Pelcovits \cite{billeter.jl:2000.a}, who used systems
of 2048 molecules modelled by the Gay-Berne potential.  The present study
investigates the effect of changing the droplet size, and uses much larger
numbers of molecules, up to one million.  This turns out to be essential to
avoid finite-size effects which are particularly important for the satellite
defect, since it causes long-range director deformation.  As well as observing
both the Saturn-ring and satellite defects, with the latter being stable only
for larger particle sizes as expected, we also observe a transition between
these two structures involving an intermediate, non-equatorial, {\em
surface-ring} defect.

The paper is organized as follows. In Sec.~\ref{sec:simulation} we 
present the computational details and molecular models we use to
simulate the liquid crystal mesophase and the interaction of the molecules
with the droplet surface. Section \ref{sec:results} contains the 
results of the simulations: density, director, order parameter,
and biaxiality profiles for the Saturn ring, satellite, and
surface ring defects.  We also compare our results to the known 
predictions of the elastic theory.
Concluding remarks are given in Sec.~\ref{sec:conclusions}. 

\section{Molecular model and simulation methods}
\label{sec:simulation}
Molecular dynamics simulations were carried out using axially symmetric
molecules interacting through the pair potential
\begin{equation}
\label{eqn:pot}
v_{ij} = \left\{\begin{array}{lr}
4 \varepsilon_0\left(\varrho_{ij}^{-12}-\varrho_{ij}^{-6} \right) + 
 \varepsilon_0, & \varrho_{ij}^6 < 2 \\
0 & \varrho_{ij}^6 > 2
\end{array} \right. \:.
\end{equation}
Here $\varrho_{ij} = (r_{ij}-\sigma_{ij}+\sigma_0)/\sigma_0$; $r_{ij}$ is the
center-center separation, $\sigma_0$ a size parameter, $\varepsilon_0$ an energy
parameter (both taken to be unity) and the orientation-dependent diameter
$\sigma_{ij}$ is defined by
\begin{widetext}
\begin{eqnarray}
\sigma_{ij}=\sigma_0
\left\{1 - \frac{\chi}{2}\left[
\frac{(\hat{\bf r}_{ij}\cdot\hat{\bf u}_i 
     + \hat{\bf r}_{ij}\cdot\hat{\bf u}_j)^2} 
     {1 + \chi(\hat{\bf u}_i\cdot\hat{\bf u}_j)}
+
\frac{(\hat{\bf r}_{ij}\cdot\hat{\bf u}_i 
     - \hat{\bf r}_{ij}\cdot\hat{\bf u}_j)^2}
     {1 - \chi(\hat{\bf u}_i\cdot\hat{\bf u}_j)}
\right]
\right\}^{-1/2}
\end{eqnarray}
\end{widetext}
where $\chi=(\kappa^2-1)/(\kappa^2+1)$, $\kappa$ being the elongation.  In this
work we used $\kappa=3$ throughout. The orientation dependence is written in
terms of the direction of the center-center vector $\hat{\bf r}_{ij} = {\bf
r}_{ij}/r_{ij}$ and the unit vectors $\hat{\bf u}_i$, $\hat{\bf u}_j$ which
specify the molecular symmetry axes.  This is a soft repulsive potential,
describing (approximately) ellipsoidal molecules; it may be thought of as a
variant of the standard Gay-Berne potential \cite{berne.bj:1972.a,gay.jg:1981.a}
with exponents $\mu = 0, \nu = 0$.  The molecular mass $m$ was taken to be
unity, and the molecular moment of inertia fixed as $I=2.5m\sigma_0^2$.

The interaction of molecule $i$ with the droplet
was given by a shifted Lennard-Jones repulsion potential between the centers,
having exactly the same form as eqn~(\ref{eqn:pot}), but with $\varrho_{ij}$
replaced by $\varrho_{i} = (|{\bf r}_{i}-{\bf r}_c|-\sigma_{c}+\sigma_0)/
\sigma_0$ and $\sigma_c = R + \sigma_0/2$. 
Here, $R$ is the colloid radius, and ${\bf r}_c$
the position of the colloid center.  This interaction potential results in
homeotropic anchoring of the liquid crystal molecules, normal to the particle
surface.

\begin{center}
\begin{table}
\caption[Parameters]{
\label{tab:1}
Run parameters}
\begin{ruledtabular}
\begin{tabular}{crcr}
$R$ & $N$ & Defect type & Steps\\
\hline
3  & 8,000      & Saturn ring               & 1,500,000 \\
5  & 8,000      & Saturn ring               & 1,500,000 \\
6  & 8,000      & Saturn ring               & 1,500,000 \\
7  & 8,000      & Saturn ring               & 1,500,000 \\
10 & 64,000    & Saturn ring               & 500,000   \\
10 & 64,000    & off-center ring           & 1,000,000 \\
15 & 1,000,000 & Saturn ring               & 500,000   \\
15 & 1,000,000 & satellite                 & 500,000   
\end{tabular}
\end{ruledtabular}
\end{table}
\end{center}

The systems consisted of $N = $ 8,000--1,000,000 particles depending on the
droplet radius (in the range $R/\sigma_0 = 3\ldots15$) and defect type: details
appear in Table~\ref{tab:1}. A molecular dynamics program was run on a Cray T3E
using a domain decomposition algorithm \cite{wilson.mr:1997.b}.  A reduced
temperature $k_{\rm B}T/\varepsilon_0=1$ was used throughout (for this model,
the phase behaviour is not sensitively dependent on temperature, as there are no
attractive forces).  The system size was chosen so that the number density of
the liquid crystal far from the colloid was $\rho\sigma_0^3=0.35$, which lies
well within the nematic region, with a corresponding bulk order parameter
$S\approx 0.81$.  For this system, in the reduced units defined by $\sigma_0$,
$\varepsilon_0$ and $m$, a timestep $\delta t=0.004$ was found suitable.

Cubic periodic boundary conditions were employed in all the simulations.  To
facilitate the analysis, the position of the droplet was fixed in the centre of
the simulation box, and a global constraint was applied to fix the orientation
of the director along the $z$ axis \cite{germano.preprint}.

To obtain the particular defect types of interest we have used several ways 
to prepare the initial configuration:
\begin{enumerate}
\item The radius of the colloid droplet was steadily increased from a small
  value in systems which were already in the nematic state. This takes about
  $10^4$ steps to increase the droplet radius up to $R/\sigma_0 = 10$ and about
  $10^4$ steps to equilibrate the system with the defect. This method generates
  an energetically favourable configuration of the kind which is stable for
  small colloid particles (i.e\ the ring defect).
\item Disordered, isotropic configurations containing the colloid droplet were
  compressed to the ordered, nematic, state.  This method, in principle, should
  give the lowest free-energy configurations in an unbiased way.  However, for
  large systems, equilibration from the isotropic phase is extremely
  time-consuming: systems split into small domains which then realign (coarsen).
\item For large systems, we prepared director configurations in the nematic
  state with the defect embedded in them.  To introduce the defect into the
  mesophase we used approximate analytical expressions for the director field
  ${\bf n}({\bf r})$ around the droplet, which were used to determine the
  initial molecular orientations. This configuration was diluted, i.e.  the
  molecule coordinates were scaled so that they did not overlap with each other
  and the colloid droplet. We compressed the system under low temperature ($T =
  0.1$) to the desired density. The low-temperature compression preserves the
  director structure in the system. Therefore, the original defect structure is
  still present in the system after the compression. After the compression, the
  system was equilibrated for $10^5$ steps. This method allows one to prepare
  and observe metastable configurations, i.e.\ those which are local minima of
  the system free energy.
\end{enumerate}
\section{Simulation results and discussion}
\label{sec:results}

\begin{figure}
 \includegraphics[width=3cm]{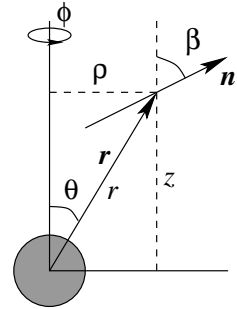}
 \caption[Coordinates]{
 \label{fig:2} 
 Coordinate systems used in this work.  The system is assumed to have rotational
 symmetry about the $z$-axis.  The director field is assumed to have axial
 ($\hat{\bf z}$) and radial ($\hat{\bf e}_\rho$) components only, and so may be
 described by the angle $\beta(r,\theta)$.  }
\end{figure}

Simulation results were analysed to give director, density, order parameter, and
biaxiality maps.  Coordinate systems used to describe positional and
orientational variables are shown in Fig.~\ref{fig:2}.  Positions, relative to
the colloid particle at the origin, are expressed in spherical polar coordinates
$(r,\theta,\phi)$; the system was assumed to be invariant to rotations about the
$z$ axis.  Accordingly, all quantities were averaged over angle $\phi$ (by
rotating molecular positions and orientations together) before accumulating
simulation averages on a grid with $\Delta r = $0.05--0.1$\sigma_0$, $\Delta
\theta = 10^\circ$.  At each grid point, the order tensor ${\bf Q}$ was resolved
into components based on cylindrical polar unit vectors, pointing (i) along the
$z$-axis $\hat{\bf z}$, (ii) in the direction $\hat{\bf e}_\rho$ of increasing
distance $\rho=\sqrt{x^2+y^2}$ from the $z$-axis, and (iii) in the tangential
direction $\hat{\bf e}_\phi=\hat{\bf z}\times\hat{\bf e}_\rho$:
\begin{widetext}
\begin{equation}
Q_{\alpha\beta}(r_i,\theta_j) =  \frac{1}{n_{\{i,j\}}} 
\sum_{k=1}^{n_{\{i,j\}}}
\left\{\frac{3}{2} \left\langle u_{k\alpha} u_{k\beta} \right\rangle
- \frac{1}{2}\delta_{\alpha\beta} \right\}
\qquad \alpha,\beta=\rho,\phi,z
\end{equation}
\end{widetext}
where there are $n_{\{i,j\}}$ molecules present in the bin $\{i,j\}$,
$\delta_{\alpha\beta}$ is the Kronecker delta, $\langle\ldots\rangle$ denotes an
ensemble average.  Diagonalizing this tensor, for each bin, gives three
eigenvalues and three corresponding eigenvectors.  The largest eigenvalue
defines the order parameter $S(r,\theta)$ for each bin.  The biaxiality $\alpha$
is defined as
\begin{equation}
\alpha = \frac{1}{3}\left(S_2 - S_3\right), 
\end{equation}
where $S_2$, $S_3$ are the two remaining eigenvalues of the averaged order
tensor ${\bf Q}$ \cite{biscari.p:1997.a}.

\subsection{Ring defect}
\label{subsec:ring}
For all studied radii ($R/\sigma_0=$ 3--15) the ring defect appears immediately
after equilibration of the system starting from the isotropic state or 
blowing up the droplet in the nematic state. We therefore concluded
that this type of the defect is energetically more favourable for the chosen
droplet sizes.
For large configurations, we also used the method with the predefined director 
field. In the case of the ring defect, we used the trial function for the 
director angle (see Fig.~\ref{fig:2})
\begin{displaymath}
\beta(r,\theta) = \arccos[n_z(r,\theta)]
\end{displaymath}
proposed by Kuksenok {\it et al}
\cite{ruhwandl.rw:1997.b,shiyanovskii.sv:1998.a}.  This trial function captures
the far-field behavior and a ring of $(-1/2)$ singularity at $r = a_{\rm r}$,
$\theta = \pi/2$:
\begin{equation}
\beta(r,\theta) = \theta - \frac{1}{2}\arctan
\frac{\sin 2\theta}{1/f(r)+\cos 2\theta}.
\end{equation}
Here $f(r)$ is a unique function of $r$ and independent of the angle $\theta$.
$f(r)$ has also to comply with the boundary conditions: $1/f(\infty) = 0$
($\Rightarrow~\beta = 0$) far from the particle and $f(R) = 0$
($\Rightarrow~\beta = \theta$) on the particle surface.The explicit form of the
function $f(r)$ can be found in 
refs~\cite{ruhwandl.rw:1997.b,shiyanovskii.sv:1998.a}.

\begin{figure}
 \includegraphics[width=6cm]{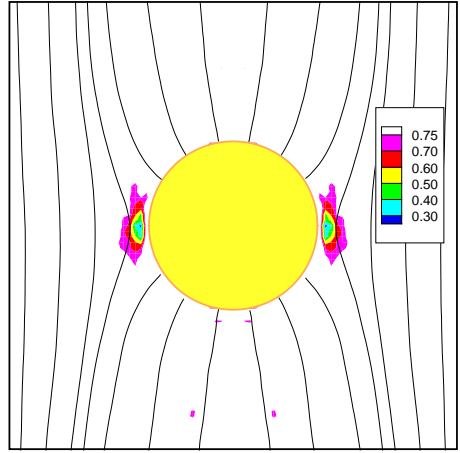}
 \caption[Ring director]{
 \label{fig:3} 
 Director map of the ring defect.  Droplet radius $R = 15$. The
 shading represents the value of the order parameter. The
 defect core is located close to the droplet surface, $a_{\rm r} \approx
 1.164R$.  The director variation rapidly vanishes in the cell bulk. }
\end{figure}

\begin{figure}
 \includegraphics[width=8cm]{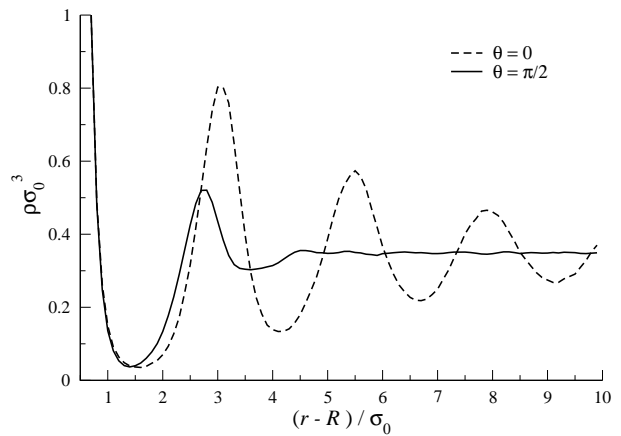}
\caption[Ring density]{
\label{fig:4} 
Density profiles for the ring defect.  We plot dimensionless number density
$\rho\sigma_0^3$ versus reduced distance $r/\sigma_0$. Droplet radius
$R/\sigma_0=15$. The following directions are shown: $\theta = 0, \pi$ (avoiding
the defect) and $\theta = \pi/2$ (crossing the disclination ring). The inset
shows the contour plot of the density map.}
\end{figure}

\begin{figure}
 \includegraphics[width=8cm]{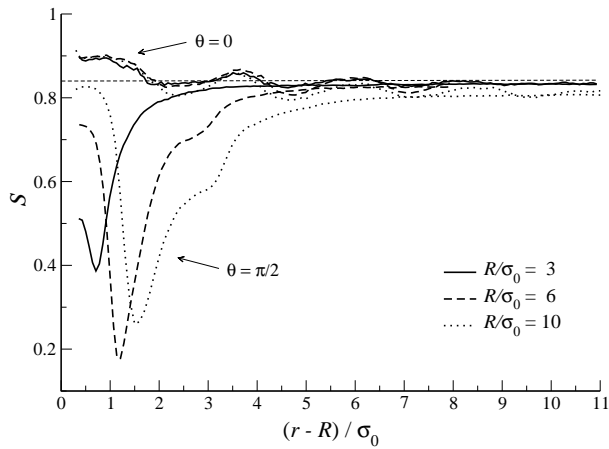}
\caption[Ring order]{
\label{fig:5} 
Order parameter profiles for the ring defect along the directions $\theta = 0,
\pi/2$.  Different curves correspond to the different droplet radii. The minimum
of the order parameter defines the position of the disclination core.  }
\end{figure}

\begin{figure}
 \includegraphics[width=8cm]{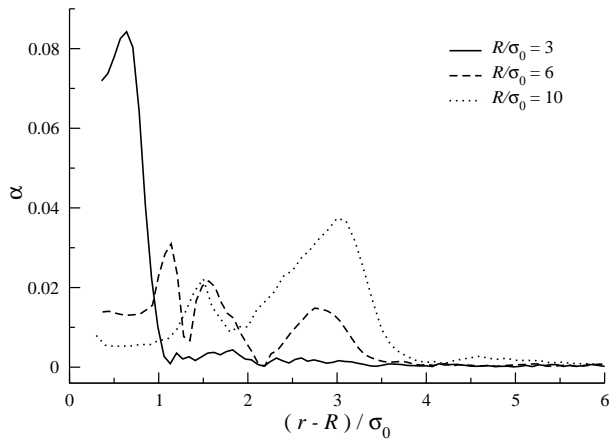}
\caption[Ring biaxiality]{
\label{fig:6} 
Biaxiality profiles along the direction $\theta = \pi/2$ (across the defect).
We plot the biaxiality parameter $\alpha$ versus reduced distance $r/\sigma_0$.
The peak of the biaxiality is centered on the defect core and coincides with the
minimum of the order parameter (Fig~\ref{fig:5}).  }
\end{figure}

\begin{figure}
 \includegraphics[width=8cm]{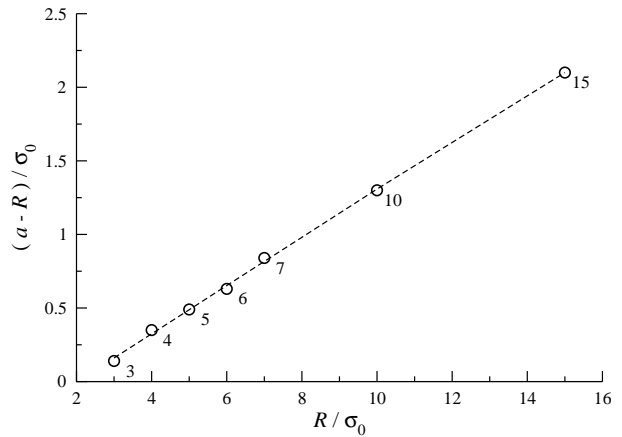}
\caption[Ring distance]{
\label{fig:7} 
Distance of the core region of the ring defect from the droplet center, as a
function of reduced droplet radius $R/\sigma_0$.  We also show the linear fit to
the simulation results, $a_{\rm r}=1.164R-0.33\sigma_0$.  }
\end{figure}

A typical director map of the ring defect obtained from the simulations is shown
in Fig.~\ref{fig:3}. From this map one can see that the ring defect does not
have long-range director distortions. Its core region is located very close to
the droplet surface and the director distortion vanishes very quickly in the
liquid crystal bulk.  This is in agreement with the {\em quadrupolar} nature of
the defect: far from the particle the director deviation angle has asymptotic
behaviour $\beta \sim (R/r)^3\sin2\theta$.  Therefore, this type of defect can
be studied using a comparatively small number of particles and correspondingly
small simulation boxes (see Table~\ref{tab:1} for details).

Typical density profiles for the ring defect with $R=15$ are shown in
Fig.~\ref{fig:4} for $\theta = 0, \pi$ (not intersecting the defect) and $\theta
= \pi/2$ (crossing the disclination ring). The profiles which avoid the
disclination have oscillating structure near the particle surface which is
typical for a {\em liquid-wall} interface. The profiles which cross the
disclination ring do not have oscillations. The difference may be due to the
partial melting of the liquid crystal in the disclination core region. This
melting damps the influence of the droplet surface on the interface region.
Note that, even though the disclination core is located close to the particle
surface, the region where the density has an oscillating structure is quite big.
Therefore, the simulation box has to be of adequate size in order not to affect
the defect structure.

The order parameter profiles for $\theta = 0, \pi/2$ are shown in
Fig~\ref{fig:5}. The shape of these profiles in general reflects the typical
structure of the core: the center of the core has lower order than the bulk and
the core region extends over a few molecular lengths.  Using order parameter
profiles one can also define the position of the disclination ring as the
location of the minimum of the order parameter.  For large enough $R$ a second
minimum appears, indicating that the disclination ring defect has a complex
structure. This minimum is probably due to the density modulation near the
droplet surface. As we will see, the density modulation also influences the rest
of the profiles.

To emphasise the complex structure of the defect core we plot the biaxiality
profiles (Fig~\ref{fig:6}).  From the order and biaxiality profiles one can see
that the main biaxial ring is accompanied by one additional biaxial ring for
large $R$, or even two extra rings for some $R$ (e.g. $R/\sigma_0 = 5,6$).
Again, this is most likely due to the density modulation near the droplet
surface.

Using the minima of the order profiles (or, with the same results, the maxima of
the biaxiality profiles) we extracted the distance from the core of the
disclination ring to the particle surface. The dependence of this distance on
the particle radius $R$ is shown in Fig~\ref{fig:7}.  It is interesting to
compare with the phenomenological theory predicting linear dependence of the
ring defect radius on the droplet radius: $a_{\rm r} \approx 1.25R$ from
minimization of the elastic free energy using a trial function
\cite{kuksenok.ov:1996.a}, $a_{\rm r} \approx 1.13R$ using a simulated annealing
method \cite{ruhwandl.rw:1997.b}, or $a_{\rm r} \approx 1.26R$
\cite{stark.h:1999}.  Our MD simulation results give $a_{\rm r} - R =
-0.33\sigma_0 + (0.164 \pm 0.004)R$ which is in good agreement with the
phenomenological theory, especially if one bears in mind the complex structure of the defect core.
\subsection{Satellite defect}
\begin{figure}
 \includegraphics[width=6cm]{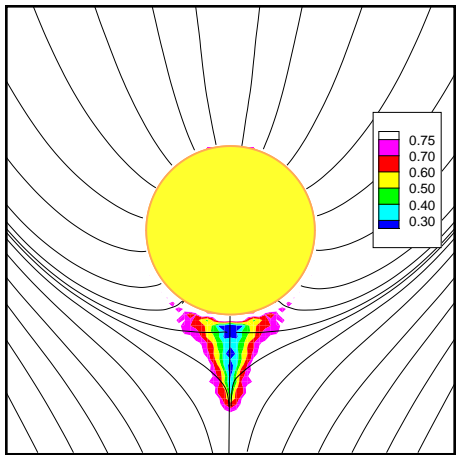}
\caption[Satellite director]{
\label{fig:8} 
Director map of the satellite defect. Droplet radius $R/\sigma_0 = 15$. 
The background contour plot represents the value of the order parameter. }
\end{figure}

The director field around the satellite defect is illustrated in
Fig~\ref{fig:8}.  The director distortion extends much further than that of the
ring defect.  This is a direct consequence of the symmetry of the director
distribution of the satellite defect: far from the particle, the director angle
vanishes as $\beta \sim (R/r)^2\sin \theta$, i.e.\ it is like a dipolar term in
a multipole expansion.  Therefore, to study the satellite defect, one needs very
large systems.  We used one million particles and droplet radius $R = 15$.

\begin{figure}
 \includegraphics[width=8cm]{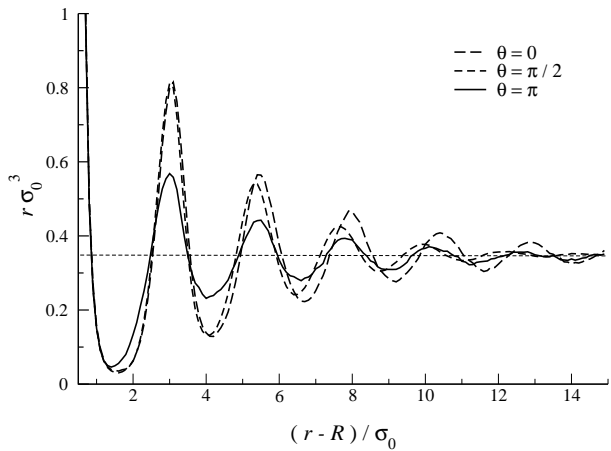}
\caption[Satellite density]{
\label{fig:9} 
\small Density profiles for the satellite defect. Droplet radius
$R/\sigma_0=15$.  The following directions are shown: $\theta = 0, \pi/2$
(both avoiding the defect) and $\theta = \pi$ (crossing the defect). 
The contour plot of the density map is shown in the inset. }
\end{figure}

\begin{figure}
 \includegraphics[width=8cm]{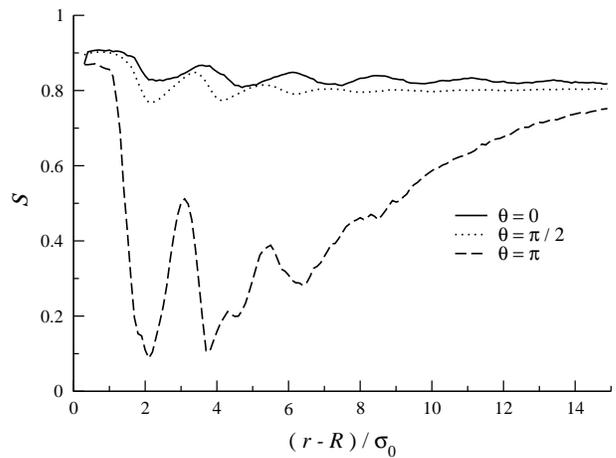}
\caption[Satellite order]{
\label{fig:10} 
Order parameter profiles for the satellite defect along the directions $\theta =
0, \pi/2, \pi$.  Droplet radius $R/\sigma_0=15$. The defect core is elongated
with the long axis parallel to the far-field director.  }
\end{figure}

\begin{figure}
 \includegraphics[width=8cm]{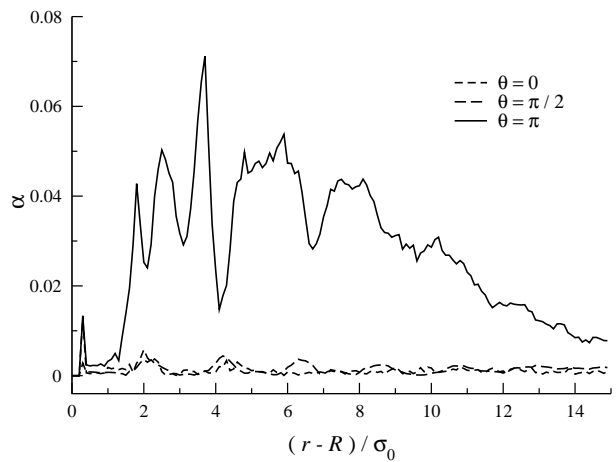}
\caption[Satellite biaxiality]{
\label{fig:11} 
Biaxiality profiles for the satellite defect ($\theta = 0, \pi/2, \pi$). Droplet
radius $R=15$. From these profiles one can define the center of the satellite
defect as a maximum of the biaxiality. The small blip close to the droplet
surface is due to the surface effects: it is seen on all profiles.  }
\end{figure}

The approach involving a trial function was used to generate the dipolar
initial configuration, as in the case of the Saturn ring.  In this case
superposition of the radial and hyperbolic hedgehogs implies the following
ansatz \cite{shiyanovskii.sv:1998.a}:
\begin{equation}
\beta(r,\theta) = \theta - \arctan
\frac{\sin \theta}{1/f(r)+\cos \theta}.
\end{equation}
Again, $f(r)$ is supposed to be a unique function of $r$ which does not depend
on the angle $\theta$. The boundary condition now is $1/f(\infty) = 0$
($\Rightarrow~\beta = 0$) far from the particle; the position of the hyperbolic
hedgehog is defined by $f(a_{\rm s}) = 1$.  After the initial configuration was
generated the system was equilibrated for $50,000$ steps with the subsequent
production runs extending to $500,000$ steps.

The density profiles for the satellite defect are shown in Fig~\ref{fig:9}.  The
following directions are shown: $\theta = \pi$ (through the defect) and $\theta
= 0, \pi/2$ (avoiding defect). One can see that the density profile at $\theta =
\pi$ (across the defect core) has less prominent oscillating structure than the
other two: this is again presumably due to the partial disordering of the
mesophase in the core region.

The order parameter profiles for $\theta = 0, \pi/2, \pi$ are shown in
Fig~\ref{fig:10}. One can see that the defect core region is quite
elongated with its long axis parallel to the far-field director.  It is
difficult to judge the position of the satellite defect from the order parameter
profiles: the region occupied by the disclination core is rather large.

The biaxiality profiles for the same values of $\theta$ are shown in
Fig~\ref{fig:11}.  From these profiles one can see that the center of the defect
core is located at a distance $a_{\rm s}\approx 1.4R$.  The value predicted by
the elastic theory is about $1.22R$ for the simulated annealing calculations
\cite{ruhwandl.rw:1997.b}; $1.17R$ \cite{poulin.p:1997.a} or $1.46R$
\cite{shiyanovskii.sv:1998.a} for the free energy minimization using a trial
function.  Again, as in the case of the ring defect, both order parameter and
biaxility profiles are affected by the density oscillations near the particle
surface.
\subsection{Off-center ring}
\label{sec:surface ring}
\begin{figure}
 \includegraphics[width=6cm]{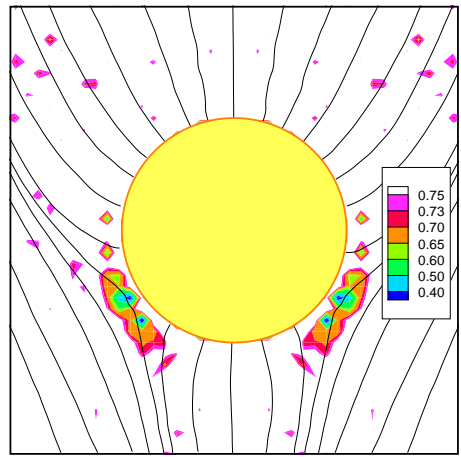}
\caption[Surface ring]{
\label{fig:12} 
Director map for the off-center ring configuration. The background contour plot 
represents the value of the order parameter.  }
\end{figure}

\begin{figure}
 \includegraphics[width=8cm]{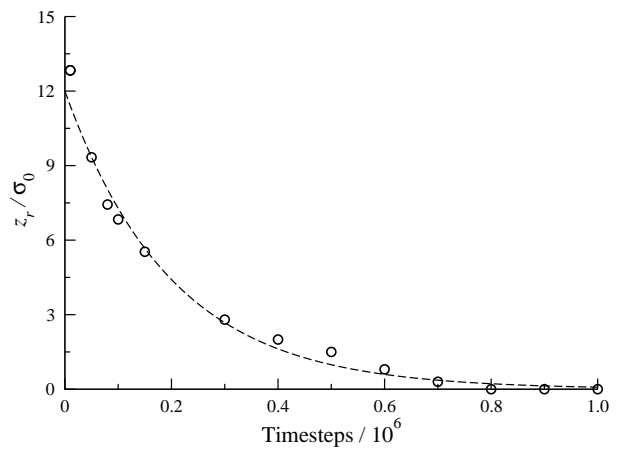}
\caption[Off-center ring position]{
\label{fig:13} 
Dynamics of the transition from the satellite defect to the Saturn ring defect 
via an off-center ring configuration. We plot the $z$ coordinate of the
ring defect relative to the droplet center, versus number of the timesteps.  As
a guide, the dashed line corresponds to exponential decay with a time constant
of $2\times10^5$ timesteps. }
\end{figure}

Simulation results show that both the satellite and ring defects are at least
metastable for $R/\sigma_0=15$: once the particular defect is realized in the
system, it is stable over the timescale that is accessible to our simulations.
However, the satellite defect is not stable for the smaller droplets.  Indeed,
we observed a rapid transition (several thousands of MD steps) of the
satellite defect to the ring defect for $R/\sigma_0 < 10$.  Equilibrating the
initial configuration with the satellite defect in the cell with droplet radius
$R/\sigma_0 = 10$, we observed that it evolves into an {\em off-centered} ring
defect: a typical nematic director map is shown in Fig~\ref{fig:12}.  The ring
moved slowly, evolving towards an equatorial Saturn ring configuration. Doing
long runs (up to a million timesteps) we conclude that this is an intermediate
state between the satellite and the Saturn ring defect. The $z$-coordinate of
the ring, relative to the droplet center, as a function of the number of the
timesteps, is shown in Fig~\ref{fig:13}. The evolution dynamics is quite slow.
In principle, one might observe the same type of transition in the opposite
direction, for droplet sizes large enough to stabilize the satellite defect;
this is still unreachable for the system sizes we explore here.  Note that
elastic theory also predicts the off-center ring configuration to be unstable
\cite{stark.h:1999}, with the transition from the dipole configuration to the
Saturn ring configuration occurring via this intermediate state.

\section{Conclusions}
\label{sec:conclusions}
We used molecular dynamics to study a spherical object, representing a small
colloidal particle or droplet, suspended in a nematic liquid crystal matrix
modelled with the soft ellipsoid potential. Homeotropic boundary conditions and
strong anchoring create a hedgehog director configuration on the particle
surface and in its vicinity; we have studied the defect structure around the
particle which cancels this hedgehog defect and produces a uniform director
field at large distances.  We investigated systems up to $10^6$ particles with a
domain decomposition program on a massively parallel computer, conducting runs
of up to $1.5\times10^6$ timesteps.

Our simulation results show that, in the case of a small droplet, only a Saturn
ring defect is stable; an initial configuration containing a satellite defect
evolves spontaneously into a ring.  For large enough droplets, however, both the
satellite and the ring defect are at least metastable on the timescale of the
simulations.  For intermediate sizes of colloid particle, a non-equatorial
surface ring evolves very slowly towards the Saturn ring configuration.

Using order parameter maps we are able to resolve the position of the core of
the defect; these also show that the nematic phase is strongly biaxial near the
defect core. In these systems, the defect region is quite large: in some cases
of a similar size to the colloid particle itself. Far from the particle the
director profiles are in good agreement with those predicted by elastic theory.
The elastic theory also predicts quite well the variation of defect position
with colloid particle radius.

It should be borne in mind that the colloid dimensions studied here are much
smaller than those typically investigated experimentally; indeed, it is quite
surprising that the satellite defect is observable in our simulations, and that
the off-center surface ring, which is never seen in the elastic theory, appears,
albeit as a transition state.  It is hoped that further work, aimed at
estimating free energy differences between these structures, will clarify the
situation.

\acknowledgments 
This research was supported by EPSRC grants GR/L89990 and
GR/M16023, and through INTAS grant 99-00312.  D.A. acknowledges the support of
the Overseas Research Students Grant; G.G. acknowledges the support of a British
Council Grant; M.P.A. is grateful to the Alexander von Humboldt foundation and
the Leverhulme Trust.
%

\end{document}